\documentclass[superscriptaddress, letterpaper,twocolumn,prl,showpacs, showkeys,reprint]{revtex4}
\usepackage{amsmath, amsthm, amssymb}
\usepackage[pdftex]{graphicx}
\usepackage{dcolumn} 
\usepackage{bm} 

\makeatletter 
\gdef\@ptsize{2}
\let\@currsize\normalsize 
\makeatother 
\usepackage{setspace} 
\usepackage[usenames,dvipsnames]{color}

\begin{document}
\title{Interplay of Spin-Orbit Interactions, Dimensionality, and Octahedral Rotations in Semimetallic SrIrO$_3$}

\author{Y.F. Nie}
\affiliation{Laboratory of Atomic and Solid State Physics, Department of Physics, Cornell University, Ithaca, New York 14853, USA}
\affiliation{Department of Materials Science and Engineering, Cornell University, Ithaca, New York 14853, USA}
\affiliation{National Laboratory of Solid State Microstructures and College of Engineering and Applied Sciences, Nanjing University, Nanjing, 210093, People?s  Republic of China}
\author{P.D.C. King}
\affiliation{Laboratory of Atomic and Solid State Physics, Department of Physics, Cornell University, Ithaca, New York 14853, USA}
\affiliation{Kavli Institute at Cornell for Nanoscale Science, Ithaca, New York 14853, USA}
\author{C.H. Kim}
\affiliation{Department of Applied Physics, Cornell University, Ithaca, New York 14853, USA}
\author{M. Uchida}
\author{H.I. Wei}
\author{B.D. Faeth}
\author{J.P. Ruf}
\affiliation{Laboratory of Atomic and Solid State Physics, Department of Physics, Cornell University, Ithaca, New York 14853, USA}
\author{J.P.C. Ruff}
\affiliation{CHESS, Cornell University, Ithaca, New York 14853, USA}
\author{L. Xie}
\affiliation{National Laboratory of Solid State Microstructures and College of Engineering and Applied Sciences, Nanjing University, Nanjing, 210093, P.R. China}
\affiliation{Department of Materials Science and Engineering, University of Michigan, Ann Arbor, Michigan 48109, USA}
\author{X. Pan}
\affiliation{Department of Materials Science and Engineering, University of Michigan, Ann Arbor, Michigan 48109, USA}
\author{C.J. Fennie}
\affiliation{Department of Applied \& Engineering Physics, Cornell University, Ithaca, New York 14853, USA}
\author{D.G. Schlom}  
\affiliation{Department of Materials Science and Engineering, Cornell University, Ithaca, New York 14853, USA}
\affiliation{Kavli Institute at Cornell for Nanoscale Science, Ithaca, New York 14853, USA}
\author{K.M. Shen}
\email[Author to whom correspondence should be addressed: ]{kmshen@cornell.edu}
\affiliation{Laboratory of Atomic and Solid State Physics, Department of Physics, Cornell University, Ithaca, New York 14853, USA}
\affiliation{Kavli Institute at Cornell for Nanoscale Science, Ithaca, New York 14853, USA}


\begin{abstract}

We employ reactive molecular-beam epitaxy to synthesize the metastable perovskite SrIrO$_{3}$ and utilize {\it in situ} angle-resolved photoemission to reveal its electronic structure as an exotic narrow-band semimetal. We discover remarkably narrow bands which originate from a confluence of strong spin-orbit interactions, dimensionality, and both in- and out-of-plane IrO$_6$ octahedral rotations. The partial occupation of numerous bands with strongly mixed orbital characters signals the breakdown of the single-band Mott picture that characterizes its insulating two-dimensional counterpart, Sr$_{2}$IrO$_{4}$, illustrating the power of structure-property relations for manipulating the subtle balance between spin-orbit interactions and electron-electron interactions.  
\end{abstract}

\maketitle

The combination of strong spin-orbit interactions (SOIs) with electron-electron correlations has recently been predicted to realize a variety of novel quantum states of matter, including topological Mott insulators~\cite{pesin2010, Zhang:CeLvzJTX}, quantum spin Hall, quantum anomalous Hall, and axion insulators~\cite{Xiao:2011eg,Shitade:2009he,Wan:2012du,Ruegg:2012jd}, Weyl semimetals~\cite{wan2011}, and even high temperature superconductors~\cite{Wang2011}. Although typically viewed to be disparate properties, the recent discovery that SOI can sufficiently enhance the effective role of electron correlations to stabilize a Mott-like insulating state in the quasi-two-dimensional $5d$ transition metal oxide Sr$_2$IrO$_4$~\cite{bjkim2008,kim2009} has opened a new frontier for exploring the rare interplay between these 2 degrees of freedom. Its three-dimensional perovskite analogue, SrIrO$_3$, has been theoretically proposed as a key building block for engineering topological phases at interfaces and in superlattices~\cite{Xiao:2011eg, carter2012, Lado:2013to}. In bulk, SrIrO$_{3}$ is believed to lie in close proximity to a metal-insulator transition~\cite{sjmoon2008,Liu:2013vi}, yet little is known of its electronic structure to date.

Here we reveal the momentum-resolved electronic structure of SrIrO$_{3}$ using a combination of reactive oxide molecular-beam epitaxy (MBE) and {\it in situ} angle-resolved photoemission spectroscopy (ARPES). Our measurements uncover an exotic semimetallic ground state, hosting an unusual coexistence of heavy holelike and light electronlike bands. Contrary to conventional expectations that increased coordination leads to broader bands in higher-dimensional materials~\cite{sjmoon2008}, we find that the bandwidths of SrIrO$_3$ are, instead, narrower than its insulating two-dimensional counterpart. By combining first-principles calculations with spectroscopic measurements, we uncover the surprising interplay of spin-orbit interactions, dimensionality, and octahedral rotations which drives the narrow-band, semimetallic state in SrIrO$_{3}$. Our results indicate that subtle changes in the structure and rotation angles should drive substantial changes in the electronic structure and physical properties of SrIrO$_{3}$. This highlights the important structure-property relationships in correlated quantum materials~\cite{PhysRevLett.106.016405,pavarini2005chemistry}, much like in ferroelectrics and multiferroics~\cite{benedek2013there}. This Letter also underscores that simple toy models which neglect these structural degrees of freedom may fail to capture the essential underlying physics of the iridates.

\begin{figure*}
\begin{center}
\includegraphics{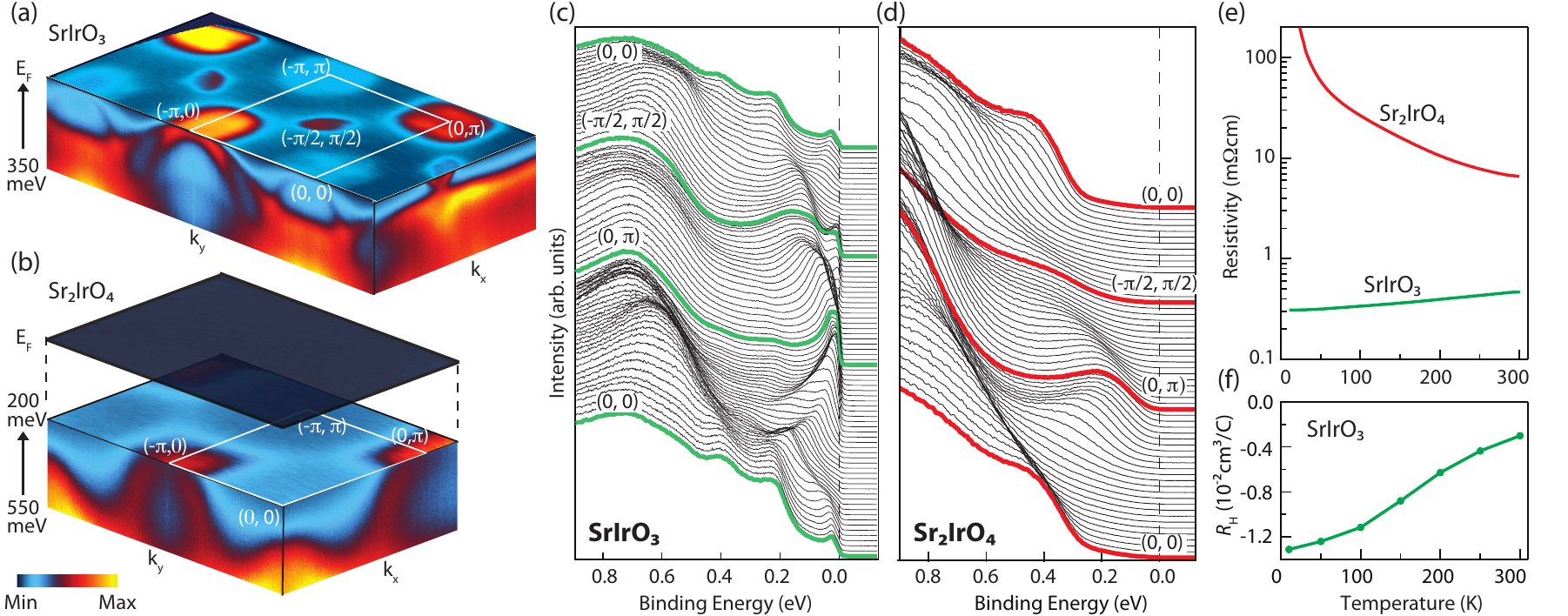}
\caption{(a, b), $E$ vs.\ $k$ dispersions and isoenergy  intensity maps of SrIrO$_3$ and Sr$_2$IrO$_4$ measured by ARPES at temperatures of 20 and 70~K, respectively. The isoenergy intensity maps are unsymmetrized, while the $E$ vs\ $k$ data have been symmetrized about $k_x=\pi/a$ for purposes of illustration. The white solid lines denote a quarter of the pseudocubic and tetragonal projected 2D Brillouin zones (2DBZs) for SrIrO$_3$ and Sr$_2$IrO$_4$, respectively. (c), (d) Corresponding energy distribution curves (EDCs) along the high-symmetry directions of the 2DBZ, revealing sharp quasiparticle peaks and extremely narrow bandwidths in SrIrO$_{3}$, compared to the broader bandwidths and wider spectral features in Sr$_{2}$IrO$_{4}$. (e), Longitudinal in-plane resistivity showing the metallic and insulating transport characteristics of SrIrO$_3$ and Sr$_2$IrO$_4$, respectively. (f), The Hall resistivity of SrIrO$_{3}$ exhibits a large temperature dependence, consistent with the complex multiband electronic structure revealed by ARPES. 
\label{fig:FSs}}
\end{center}
\end{figure*}

The perovskite SrIrO$_{3}$ is metastable in bulk~\cite{Longo1971,Zhao:2008hh}. SrIrO$_{3}$ and Sr$_{2}$IrO$_{4}$ films were synthesized by MBE with an oxidant (distilled O$_{3}$) background pressure of $1\times10^{-6}$~Torr and at a substrate [(001) (LaAlO$_3$)$_{0.3}$ (SrAl$_{1/2}$Ta$_{1/2}$O$_3$)$_{0.7}$ (LSAT)] temperature of 700$^{\circ}$C~\cite{supplemental}.  Following MBE growth, the samples were transferred through ultrahigh vacuum ($9\times10^{-11}$~Torr) into a high-resolution ARPES system with a VG Scienta R4000 electron analyzer and a VUV5000 helium plasma discharge lamp. Measurements were performed using He~I$\alpha$ ($h\nu=21.2$~eV) photons with an energy resolution $\Delta E$ of 8~meV. Lab-based x-ray diffraction (XRD)  measurements using Rigaku Smartlab confirmed the correct pseudocubic perovskite and tetragonal structure with the out-of-plane lattice constants of 4.02(1) {\AA} and 12.90(1) {\AA} for SrIrO$_3$ and Sr$_2$IrO$_4$ films, respectively. More detailed synchrotron-based XRD measurements performed at CHESS and selected area electron diffraction measurements show no relaxation of the SrIrO$_3$ film and it has the same Pnma space group and rotation pattern ($a^-b^+a^-$, Glazer notation) as that in the bulk~\cite{Zhao:2008hh} with the $b$ axis along the [100] (equivalently [010]) direction of the LSAT substrate (Supplemental Material Fig. S2~\cite{supplemental}). Longitudinal and Hall resistivity were measured with a standard four-point techniques in a Physical Property Measurement System (PPMS) from Quantum Design. We carried out density functional theory calculations within the local-density approximation (LDA) including SOI using OPENMx \cite{OPENMx, codes}, based on the linear-combination-of-pseudo-atomic-orbitals method~\cite{ozaki}. For the calculations shown here, we have fixed the structure to the thin films measured. We find only relatively minor qualitative differences if we perform the calculations for the bulk structure from Ref.~\cite{Zhao:2008hh} (Supplemental Material Fig.~S4~\cite{supplemental}).

Figure~\ref{fig:FSs} compares the electronic structure of SrIrO$_3$ and Sr$_2$IrO$_4$ epitaxial films. Insulating Sr$_{2}$IrO$_{4}$ exhibits negligible spectral weight at the Fermi level ($E_F$), with a broad peak at $(\pm\pi,0)$ [$(0,\pm\pi)$] of the surface projected tetragonal Brillouin zone marking the top of the valence band $\sim\!200$~meV below $E_F$, consistent with earlier measurements on single crystals~\cite{bjkim2008, QWang2012,King2013}. In contrast, we find multiple narrow bands within $200$~meV of $E_F$ in SrIrO$_3$, with sharp quasiparticle peaks and a well defined Fermi surface comprising holelike pockets around $(\pm\pi, 0)$ [$(0,\pm\pi)$] and $(0,0)$ as well as elliptical electronlike pockets at $(\pm\pi/2,\pm\pi/2)$ in the surface projected pseudocubic Brillouin zone. This semimetallic electronic structure is consistent with our electrical transport measurements [Figs.~\ref{fig:FSs}(e) and \ref{fig:FSs}(f))]: The resistivity of SrIrO$_3$ samples exhibit metallic temperature behavior with no signs of weak localization, while Hall effect measurements show a strong temperature dependence, indicative of the presence of multiple types of carriers. The absolute value of the Hall coefficient is much larger than in typical metals, indicating small carrier densities or carrier compensation of electrons and holes. In a single-band model, the measured Hall coefficient at 10~K of $R_H=-0.013$~cm$^3$/C corresponds to only 0.029 electrons per Ir atom, consistent with another recent report~\cite{Liu:2013vi}.  

\begin{figure}
\begin{center}
\includegraphics{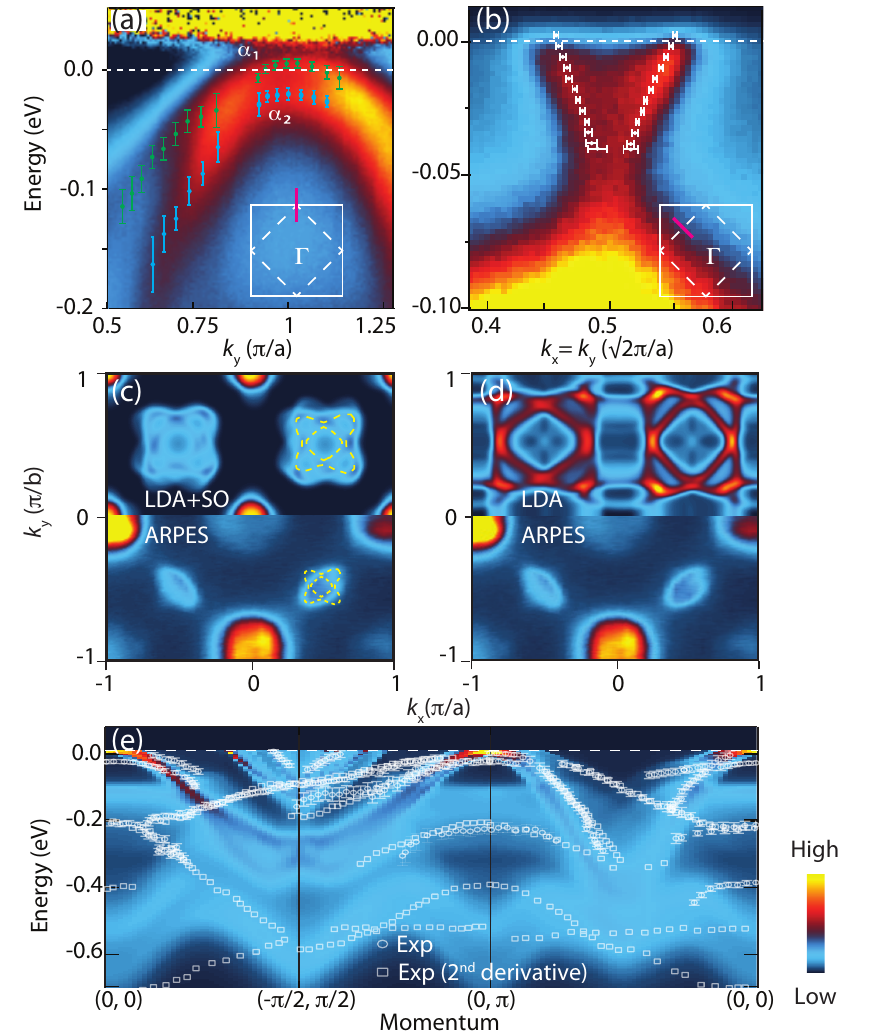}
\caption{ARPES measurements of (a) two massive holelike bands  at $(0,\pi)$, and (b) a light, almost linearly dispersive electron band at $(-\pi/2,\pi/2)$ along the cuts shown by red lines in the insets. In (a), the spectrum was measured at a temperature of 40~K and divided by the Fermi function to thermally populate states slightly above $E_F$. Lorentzian peak fits to (a) EDCs and (b) momentum distribution curves (MDCs) are shown as the symbols. (c)--(d) Simulated Fermi surfaces from the LDA+SOI calculations, assuming an inner potential of 11~eV as discussed in the main text, reproduce the general features of the measured Fermi surfaces if spin-orbit interactions are included (c), but show substantial qualitative discrepancies if they are neglected (d). The dashed lines are guides to the eye, showing four-lobe shape pockets. More details in Supplemental Material Fig.~S3.~\cite{supplemental}. (e) Simulated ARPES spectrum from the LDA+SOI calculations show good qualitative agreement with experimental measurements (symbols).}
\label{fig:LDA}
\end{center}
\end{figure}

Supporting this, our ARPES measurements reveal that the electron- and holelike bands that comprise the Fermi surface of SrIrO$_3$ have very different effective masses. For example, around $(\pm\pi,0)$ [$(0,\pm\pi)$], we find two closelyspaced holelike bands, labeled as $\alpha_1$ and $\alpha_2$ in Fig.~\ref{fig:LDA}(a), with heavy quasiparticle masses of approximately $2.4\pm0.5$ and $6\pm2$ electron masses, respectively. By thermally populating states above $E_F$, we locate the top of the upper $\alpha_1$ band at just $3\pm2$~meV above the Fermi level. In contrast, the electronlike bands around $(\pm\pi/2,\pm\pi/2)$, shown in Fig.~\ref{fig:LDA}(b), are much lighter. They exhibit approximately linear, but anisotropic, dispersions over an extended energy range of $\sim\!50$~meV, with high Fermi velocities of $0.5\pm0.1$ and $1.2\pm0.1$~eV\AA{} along the major and minor axis of their elliptical Fermi pocket, respectively. Despite the high density of states that may be expected from the heavy hole bands at the Fermi level, the light quasiparticle masses of these electron bands allow them to dominate the transport measurements, explaining the negative sign of the measured Hall coefficient and the validity of using a single-band model. The steep linear dispersion of this electron band is suggestive of a possible Dirac cone in the bulk electronic structure, reminiscent of a symmetry-protected line node predicted by previous theoretical calculations~\cite{carter2012}.

As well as this electron band, density functional calculations reproduce the other measured ARPES dispersions surprisingly well despite the absence of a Coulomb repulsion term $U$. At a fixed photon energy, ARPES measurements correspond to an almost two-dimensional projection of the three-dimensional $k$ space, centered at a value of $k_z$ which depends on the photon energy and the inner potential~\cite{Hufner}.  For a wide range of inner potentials (8--14~eV) typically observed for perovskite transition metal oxides~\cite{Yoshida:2005ee, PhysRevLett.100.056401,PhysRevB.72.054444}, the projected calculations show qualitative agreement with our ARPES measurements. We find the best agreement with theoretical calculations taking a value of 11~eV. This corresponds to a slice centered at $k_z=$ $\sim0.5~\pi/c$, where $c=4.02$ \AA{} is the pseudocubic $c$-axis lattice constant of our film. Incorporating a finite $k_z$ broadening described by a Lorentzian function with $\Delta k_{z} \approx 0.3~\pi/c$ due to the surface sensitivity of photoemission and an energy broadening through an imaginary self energy, $\Sigma''(\omega)$ = 0.1$E$$_\mathrm{binding}$, we simulate the Fermi surface spectrum from theoretical calculations [Fig.~\ref{fig:LDA}(c)].  We find multiple small pockets making up the Fermi surface. These are holelike around $(0,0)$ and $(\pm\pi,0)$ [$(0, \pm\pi)$] and electronlike around $(\pm\pi/2,\pm\pi/2)$ of the surface Brillouin zone, in good agreement with measurements, although with some small quantitative differences which could be due to electron correlations and other detailed factors.
Detailed analysis indicates that the electronlike pockets around $(\pm\pi/2,\pm\pi/2)$ actually consist of four lobes, two of which exhibit only weak spectral weight (Supplemental Material Fig.~S3~\cite{supplemental}). If we neglect SOI, however, we find qualitative disagreement with experimental measurements (Fig.~\ref{fig:LDA}(d)). These calculations predict a hypothetical Fermi surface composed of large sheets which enclose a significant portion of the Brillouin zone, touching at the extrema of the electron and hole pockets. Clearly, the opening of large hybridization gaps around these band crossings by the strong SOI has a major influence on the electronic structure, fragmenting the Fermi surface, underscoring the importance of SOI in this compound. Our LDA+SOI calculations reproduce the narrow experimental bandwidths within a factor of 1--2 [Fig.~\ref{fig:LDA}(e)],  indicating that electron correlations, which typically further narrow the band width, play a relatively minor role in SrIrO$_3$.

\begin{figure}
\begin{center}
\includegraphics {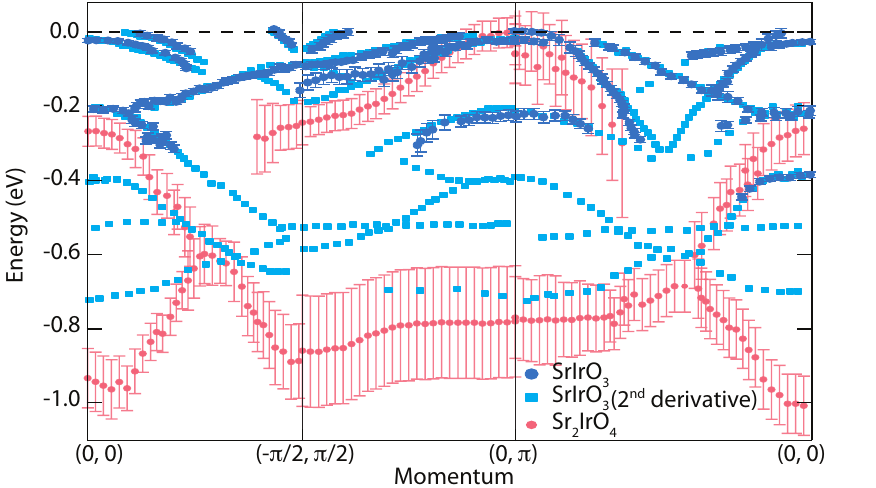}
\caption{Extracted band dispersions of SrIrO$_3$ (blue) and Sr$_{2}$IrO$_{4}$ (red) are determined by fitting both EDCs and MDCs in the raw data (circles) and estimated from peak positions in second derivative data (squares). The band dispersion of Sr$_{2}$IrO$_{4}$ is rigidly shifted in energy by 0.17 eV to align the top of the valence band  with the Fermi level in SrIrO$_3$ for purposes of comparison. \label{fig:Bandwidth}}
\end{center}
\end{figure}

Remarkably, the bandwidths of SrIrO$_3$ extracted experimentally are extremely narrow, as shown in Fig.~\ref{fig:Bandwidth}. They are on the order of only 0.3~eV estimated from multiple bands within $\sim\!1$~eV of the Fermi level, which are more typical of values found in $3d$ transition metal oxides such as cuprates or manganites that have much stronger electron-electron interactions~\cite{RevModPhys.75.473} than 5$d$ transition metal oxides. Surprisingly, they are even narrower than in the Mott-like insulating state of the single-layer compound, Sr$_2$IrO$_4$, where the bandwidths range from $\sim\!0.3 -- 0.8$~eV (Fig.~\ref{fig:Bandwidth}) as shown in our ARPES measurements and earlier measurements on single crystals~\cite{bjkim2008, QWang2012,King2013}. This is in striking contrast to the conventional pictures of increasing dimensionality, where electronic bands are expected to broaden with increased coordination.

\begin{figure}
\begin{center}
\includegraphics{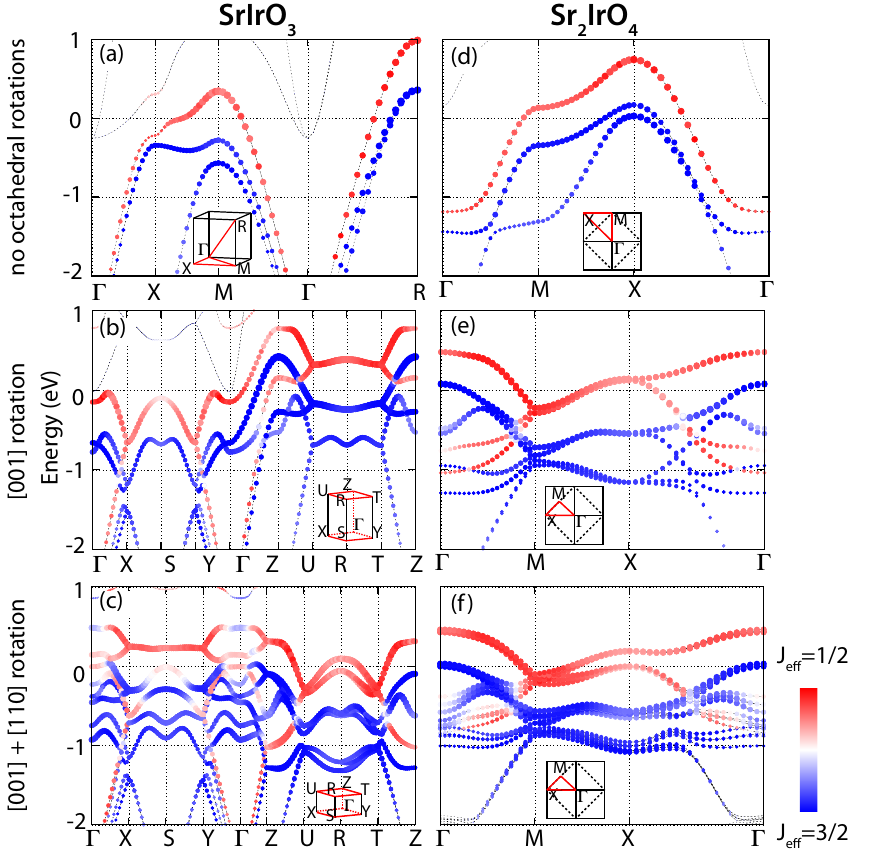}
\caption{(a), LDA+SOI band structure, projected onto a $J_\mathrm{eff}=(1/2,3/2)$ basis, of SrIrO$_3$ for a perfect cubic structure, revealing broad bandwidths. (b) Including [001] octahedral rotations backfolds the bands in plane, but leads to only relatively minor changes in the degree of orbital mixing. (c) Additionally including [110] rotations, however, causes substantial further band folding along $k_z$, leading to the intersection of numerous bands of different orbital character. SOI opens hybridization gaps around the resulting band crossings, substantially narrowing the bandwidth, and leads to strong mixing of $J_\mathrm{eff}$ = (1/2, 3/2) states. The calculations are shown along the $k$ paths (red) shown in inset. (d)--(f), Equivalent calculations of two-dimensional Sr$_2$IrO$_4$ show moderate band narrowing due to [001] rotations and only small changes due to additional artificial [110] rotations. }
 \label{fig:Breakdown}
\end{center}
\end{figure}

We attribute this to a cooperative interaction of the SOI, dimensionality, and complex octahedral rotation patterns. Considering first a hypothetical cubic structure of SrIrO$_3$ [Fig.~\ref{fig:Breakdown}(a)], we find broad bandwidths as expected for a $5d$ system. Consistent with the increased coordination, cubic SrIrO$_3$ exhibits larger bandwidths than Sr$_2$IrO$_4$ without structural distortions (Fig.~\ref{fig:Breakdown}(d))~\cite{sjmoon2008}. Including octahedral rotations about the $[001]$ direction, experimentally observed in both compounds, the calculations show a backfolding of the electronic bands in the surface plane. The experimental structure of bulk SrIrO$_3$~\cite{Zhao:2008hh} and our films (Supplemental Material Fig.~S2~\cite{supplemental}), however, are also characterized by large additional out-of-plane octahedral rotations along the $[110]$ pseudocubic direction. Combined with the significant $k_z$ dispersion of the electronic structure in this three-dimensional compound, these lead to additional band folding along $k_z$, giving rise to a significantly larger number of intersecting bands than are produced by $[001]$ rotations alone. SOI opens small energy gaps at the new band crossings, substantially narrowing the bandwidths [Figs.~\ref{fig:Breakdown}(b) and \ref{fig:Breakdown}(e)]. In contrast, no such $[110]$ rotations are found experimentally in Sr$_2$IrO$_4$. Because of the lack of Ir-O-Ir bonds along the $c$ axis in the 2D compound, however, we find that the bandwidths are not substantially affected even after artificially applying these $[110]$ rotations [Fig.~\ref{fig:Breakdown}(f)]~\cite{cationdisplacements}. Our calculations, therefore, reveal how the combination of SOI,  dimensionality, and both in- and, crucially, out-of-plane octahedral rotations lead to the array of narrow intersecting bands in SrIrO$_3$ as we observe experimentally.

The orbital characters and orbital mixing of the $t_{2g}$ bands are also substantially affected by the interplay of dimensionality and complex octahedral rotations. In SrIrO$_3$, the projected $J_\mathrm{eff}=1/2$ and 3/2 states reveal little mixing with [001] rotations [Fig.~\ref{fig:Breakdown}(b)] but much more substantial mixing with the $[110]$ rotations [Fig.~\ref{fig:Breakdown}(c)]. On the contrary, there is little mixing of $J_\mathrm{eff}=1/2, 3/2$ states in Sr$_2$IrO$_4$ even with the artificial $[110]$ rotations [Fig.~\ref{fig:Breakdown}(f)]. This is because the backfolding along $k_z$ of the quasi-two-dimensional electronic structure cannot mix bands of different orbital character but, instead, just gives rise to a weak energy splitting between pairs of bands of the same orbital character. Together, the multiband nature and extensive mixing of $J_\mathrm{eff}=1/2$ and 3/2 orbitals suggest that the single band $J_\mathrm{eff}=1/2$ limit that characterizes Sr$_2$IrO$_4$ is no longer sufficient to describe SrIrO$_3$.  

The observation of the semimetallic ground state in SrIrO$_3$ despite its ultranarrow bandwidths is striking, as the conventional Mott-Hubbard picture would anticipate a larger gap due to the narrower bandwidth~\cite{sjmoon2008}.  Instead, the numerous partially occupied bands of mixed orbital character likely place SrIrO$_3$ further away from the canonical Mott or Slater scenario typically invoked for a single, half-filled band, which may explain why SrIrO$_3$ remains metallic while Sr$_2$IrO$_4$ is insulating. Nonetheless, a more rigorous theoretical treatment of electron-electron correlations, like dynamical mean field theory, may provide in-depth understanding of this.

Our spectroscopic studies have uncovered the electronic structure of SrIrO$_{3}$, revealing a semimetallic ground state with unusually narrow bandwidths. Our work demonstrates how the electronic structure and physical properties of correlated materials can be controlled by the subtle interplay between octahedral rotations, spin-orbit interactions, and dimensionality. In particular, simple theoretical models which neglect one or more of these parameters may be unable to capture the delicate competition between these parameters. In multiferroics or ferroelectrics, small structural changes or instabilities can give rise to substantial changes in the physical properties~\cite{benedek2013there}. Here, we clearly show that in the case of SrIrO$_{3}$ and other iridates, a similar situation can arise in correlated quantum materials. This demonstrates the essential role of structure-property relations in correlated systems and opens an effective route for tuning the spin-orbit coupled ground state of interacting electron systems by small structural manipulations, for instance, through epitaxial strain, chemical, or externally applied pressure.

We gratefully acknowledge insightful discussions with K. Haule, H.Y. Kee, G. Kotliar, N. Nagaosa, T.W. Noh, N.V.C. Shen, and A. Vishwanath. This work was supported by the Air Force Office of Scientific Research (Grants No.\ FA9550-12-1-0335 and No. FA9550-11-1-0033) and the National Science Foundation (Grant No.\ DMR-0847385) through the MRSEC Program (Cornell Center for Materials Research, Grant No. DMR-1120296). This work was performed in part at the Cornell NanoScale Facility, a member of the National Nanotechnology Infrastructure Network, which is supported by the National Science Foundation (Grant No. ECCS-0335765). L. X. and X. P. acknowledge the support from the State Key Program for Basic Research of China (Grant No. 2015CB654901). M.U. acknowledges support  from  Postdoctoral Fellowship for Research Abroad (No. 24-162) from the Japanese Society for the Promotion of Science.  H.I. W. acknowledges a National Science Foundation Graduate Research Fellowship under Grant No. DGE-1144153. J.P.R. acknowledges support from the NSF IGERT Program (No. DGE-0903653). Research conducted at the Cornell High Energy Synchrotron Source (CHESS) is supported by the National Science Foundation and the National Institutes of Health/National Institute of General Medical Sciences under NSF Grant No. DMR-1332208.



\begin{thebibliography}{37}
\bibitem[{cod()}]{codes}
\bibinfo{note}{Calculations were also performed using WIEN2K, VASP, and Elk,
  which showed qualitatively similar results.}
\bibitem[{cat()}]{cationdisplacements}
\bibinfo{note}{We note that additional cation $A$-site displacements in
  SrIrO$_3$, as observed experimentally, are not included in the calculation
  shown in Fig.~\ref{fig:Breakdown}(c). However, compared to the pure rotation
  modes, we find that these have only a minor influence.}
\end{thebibliography}

\end{document}